\documentclass[twocolumn,showpacs,preprintnumbers,amsmath,amssymb,floatfix,nofootinbib]{revtex4}

\usepackage{graphicx}
\usepackage{subfigure}

\begin{document}
\input epsf
%\voffset =   0.5truein    
\draft
%\renewcommand{\topfraction}{0.8}
%----------------------------------------
%\newcommand{\picdir}[1]{./Pics/#1} %choose if pics are in Pics directory
\newcommand{\picdir}[1]{./#1}     %choose if pics are in same dir as file
\newcommand{\gpicdir}[1]{/home/felder/mypapers/hybrid/figs/#1}
%       choose if gary pics are in HIS directory
%\newcommand{\gpicdir}[1]{./Pics/#1} %choose if gary pics are in Picsdirectory
%\newcommand{\gpicdir}[1]{./#1} %choose if pics are in same dir as file
%-----------------------------------------
\def\pbg{\hbox{${\mathbf \pi\beta\gamma}$}}
\def\tsigma{\tilde{\sigma}}
\def\lesssim{\mathrel{\hbox{\rlap{\hbox{\lower4pt\hbox{$\sim$}}}\hbox{$<$}}}}
\def\gtrsim{\mathrel{\hbox{\rlap{\hbox{\lower4pt\hbox{$\sim$}}}\hbox{$>$}}}}

% Neil's Macros
\def\grad{\vec{\nabla}}
\def\lsim{\mbox{\raisebox{-.6ex}{~$\stackrel{<}{\sim}$~}}}
\def\gsim{\mbox{\raisebox{-.6ex}{~$\stackrel{>}{\sim}$~}}}

\title{Cosmological Fluctuations from Infra-Red Cascading During Inflation}

\author{Neil Barnaby$^{1}$, Zhiqi Huang$^{1}$, Lev Kofman$^{1}$ and Dmitry Pogosyan$^{1,2}$}
\address{${}^1$ CITA, University of
Toronto, 60 St.~George Street, Toronto, ON M5S 1A7, Canada}
\address{${}^2$  Physics Department,  University of Alberta, Edmonton,  Canada}

\date{February 3, 2009}

\begin{abstract} 
We propose a qualitatively new mechanism for generating cosmological fluctuations from inflation.  The non-equilibrium excitation of interacting scalar fields often evolves into infra-red (IR) and ultra-violet (UV) cascading,
resulting in an intermediate scaling regime.  We observe elements of this phenomenon in a simple model with inflaton $\phi$ and iso-inflaton $\chi$ fields interacting during inflation via the coupling $g^2 (\phi-\phi_0)^2 \chi^2$.
Iso-inflaton particles are created during inflation when they become instantaneously massless at $\phi=\phi_0$, with occupation numbers not exceeding unity.  Previous studies have focused on the momentary slowing down
of the condensate $\phi(t)$ by back-reaction effects.  Here, we point out that very quickly the produced $\chi$ particles
become heavy and their multiple re-scatterings off the homogeneous condensate $\phi(t)$ generates Bremsstrahlung radiation of light inflaton IR fluctuations with high occupation numbers.  The subsequent evolution of these IR
fluctuations is qualitatively similar to that of the usual inflationary fluctuations, but their initial amplitude is different.   The IR cascading  generates a bump-shaped contribution  to the cosmological curvature fluctuations, 
which can even  dominate over the usual fluctuations  for $g^2>0.06$.  The  IR cascading curvature fluctuations are significantly non-gaussian and the strength and location of the bump are model-dependent,  through $g^2$
and $\phi_0$.  The effect from IR cascading fluctuations  is significantly larger than that from the momentary  slowing-down of $\phi(t)$.   With a sequence of such bursts of particle production, the superposition of
the bumps can lead to a new  broad band non-gaussian component of cosmological fluctuations  added to  the usual fluctuations.  Such a sequence of particle creation events can, but need not, lead to trapped inflation.
\end{abstract}

\pacs{PACS: 98.80.Cq, CITA-2008-60,  hep-th/0902.0615}

\maketitle

%\section{Introduction}

%{\bf Introduction:} 

\section{Introduction}

In addition to the standard mechanism for generating cosmological perturbations during inflation from the vacuum fluctuations of the inflaton field \cite{fluct}, there
are alternative mechanisms including modulated fluctuations (inhomogeneous preheating) \cite{modulated,modulated2} and the curvaton \cite{curvaton}, both of which are based on the vacuum fluctuations of 
iso-inflaton fields during inflation.   In this paper we propose a new and qualitatively different mechanism for generating cosmological fluctuations during inflation.

Physical processes during inflation may leave their imprint as features in the cosmological fluctuations. These can, in principle, be observed if they fall in the 
range of the wavelengths between $10^4$ Mpc and $100$ Kpc, which corresponds to about ten e-folds during inflation.  There may also be signatures such as additional features at the horizon scale or potential observables on much smaller scales.
Relevant dynamical models were studied in the early days of the inflationary theory, for example the model with phase transitions during inflation yielding associated features in the  cosmological
fluctuations \cite{KL,KP,BBS} (see also \cite{ng}). There the time-dependent dynamics of the inflaton field $\phi$ can trigger a phase transition in the iso-inflaton $\chi$ field.  The  growth of $\chi$ inhomogeneities induces curvature fluctuations
on scales leaving the horizon at the moment of the phase transition.

Recently, several studies \cite{KL,T,B,S,E} considered features in the cosmological fluctuations from the effect of particle creation during inflation, which can be modeled by the simple interaction
\begin{equation}\label{coupling}
  \mathcal{L}_{\mathrm{int}} = -\frac{g^2}{2} (\phi-\phi_0)^2 \chi^2 
\end{equation}
with some  value of the scalar field  $\phi_0$ which the rolling $\phi(t)$ crosses 
during inflation that must be  tuned to  give a signal in the observable range of e-folds.
There are different motivations for the model (\ref{coupling}).
The early study \cite{KL99} introduced the possibility of slowing down the 
fast rolling inflaton using particle creation via the interaction (\ref{coupling}).
Imagine there are a number of field points $\phi_{0i}$, $i=1,2,\cdots, n$, where
the iso-inflaton field becomes massless and $\chi$ particles are created. The produced $\chi$ particles are diluted
by the expansion of the universe, however, the back-reaction effect from multiple bursts of particle creation may slow down the motion of $\phi$
sufficiently to allow for slow-roll inflation. 
This is called trapped inflation.  A more concrete string theory realization of trapped inflation, based on the sequence of $D3$ branes interactions, was discussed in \cite{B}.  The work \cite{E}, 
which is complimentary to this study, provides a detailed realization of trapped inflation in the context of the string theory model \cite{E0}. 

The instant of $\chi$-particle creation and the slow-down of the rolling inflaton shall generate a feature in the power spectrum of scalar curvature fluctuations from inflation
$P_{\zeta}(k)$.  This was noticed in \cite{T}, where the features in
the power spectrum were estimated from the simple-minded formula $P_{\zeta}(k)\sim \left( \frac{H^2}{\dot \phi}\right)^2$.
The inflaton slow-down was described by the mean-field equation 
\begin{equation}\label{field}
 \ddot \phi +3H\dot \phi + V_{\, ,\phi}+ g^2 (\phi-\phi_0)
 \langle \chi^2 \rangle  = 0.
\end{equation}
The vacuum expectation value (VEV) $\langle \chi^2 \rangle$ can be calculated with the analytic machinery of
particle creation with the coupling  (\ref{coupling}), which was developed in the theory of
preheating after inflation \cite{KLS,KLS97}. The QFT of $\chi$ particles interacting with the
time-depended condensate $\phi(t)$ deals with the eigenmodes $\chi_k(t) e^{i\vec{k}\cdot\vec{x}}$, where the  time-dependent
mode function obeys an oscillator-like equation in an expanding universe
\begin{equation}\label{field1}
 \ddot \chi_k +3H\dot \chi_k + \left[ \frac{{\vec k}^2}{a^2} + g^2 (\phi(t)-\phi_0)^2 \right] \chi_k=0  \ ,
\end{equation}
with time-dependent frequency $\omega_k(t)$. 
When $\phi(t)$ crosses the value $\phi_0$, the $\chi_k$ mode becomes massless and $\omega_k(t)$ varies non-adiabatically. Around 
this point $(\phi(t)-\phi_0)\approx \dot \phi_0 (t-t_0)$, where 
$t=t_0$ is corresponding time instant. With this very accurate \cite{KLS97} approximation, 
one can solve the equation (\ref{field1}) analytically to obtain the occupation number of created $\chi$ particles
\begin{equation}\label{number}
n_k= \exp{\left(-\frac{ \pi k^2}{k_\star^2} \right)} \ , \,\,\, k_\star^2=g |\dot \phi_0| \ ,
\end{equation}
presuming that $k_\star > H$. The latter condition requires coupling constant
$g > H^2/|\dot \phi_0| \sim 10^{-4}$. It is useful to note that, independent of the 
details of $V(\phi)$ and $\phi(t)$, the scale $k_\star$ can be related to the naively estimated amplitude of 
vacuum fluctuations as $k_\star/H = \sqrt{g / ( 2\pi \mathcal{P}_{\zeta}^{1/2} )}$.
Thus $k_\star/H \sim 30$ if $\mathcal{P}_{\zeta}^{1/2} = 5\times 10^{-5}$ as suggested by the CMB and the coupling is 
$g^2 \sim 0.1$.\footnote{We are assuming that supersymmetry protects the inflaton potential from radiative corrections at $t=t_0$.  An explicit realization of the type of coupling we are interested in, based on global $\mathcal{N} = 1$ supersymmetry, 
has been provided in \cite{SUSY}.  For string theory models the reader is referred to \cite{B,E,E0}.}

\begin{figure}[htbp]
\bigskip \centerline{\epsfxsize=0.4\textwidth\epsfbox{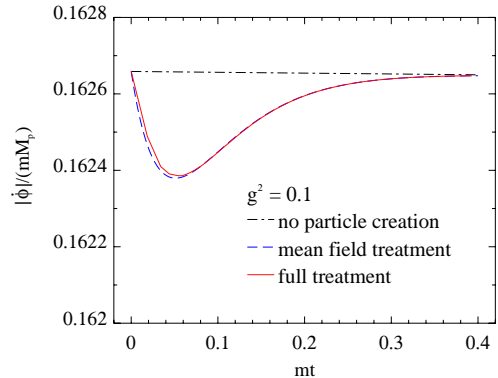}}
%\begin{verse}
%\vskip-0.25cm
\caption{$|\dot\phi|/(M_pm)$ plotted against $m t$ for $g^2=0.1$ (where $m=V_{,\phi\phi}$ is the effective inflaton mass). Time $t=0$ corresponds to the moment when $\phi=\phi_0$ and $\chi$-particles are produced copiously. 
The solid red line is the lattice field theory result taking into account the full dynamics of re-scattering and IR cascading while the dashed blue line is the result of a mean field theory treatment which ignores
re-scattering \cite{S}.  %The solid red line is the output of a lattice field theory simulation while the dashed blue line is the analytic solution obtained in a simplified treatment that only considers $\chi$-particle dilution \cite{S}. 
The dot-dashed black line is the inflationary trajectory in the absence of particle creation.
}
\label{Fig:dotphi}
\end{figure}

The VEV $\langle \chi^2 \rangle$, which controls the back-reaction on the homogeneous field $\phi(t)$,  can be calculated from (\ref{number}) and estimated as 
$\langle \chi^2 \rangle=\int \frac{d^3k}{(2\pi)^3} |\chi_k|^2 \approx  \int \frac{d^3 k}{(2\pi)^3} \frac{n_k}{\omega_k} \approx \frac{n_\chi a^{-3}}{g|\phi-\phi_0|}$ for $\phi > \phi_0$.  Substitution of this 
results back into (\ref{field}) gives expected  velocity dip of $\phi(t)$ and, correspondingly, a bump in the power spectrum $P_{\zeta}(k)$.  In Fig.~\ref{Fig:dotphi} we illustrate
this velocity dip for the model (\ref{coupling}) with $g^2=0.1$.

The calculation of curvature fluctuations in the model (\ref{coupling}) was re-considered in \cite{S}, where the linearized equations of motion for the quantum fluctuations
$\delta \phi$ coupled with the metric fluctuations were treated again in the mean-field approximation,  using $\langle \chi^2 \rangle$ to quantify 
the back-reaction. This study shows that the bump in the curvature power spectrum is the most prominent  part of an otherwise wiggling pattern. Similar to us, the work \cite{E} further refined the calculation
of the curvature perturbation in this model, going beyond the mean-field treatment of $\phi$.\footnote{Below we use QFT methods to study correlators of
inhomogeneous fluctuations $\delta \phi$ induced by $\chi^2$ inhomogeneities.  Ref.\ \cite{E} considers the effect induced by quantum mechanical fluctuations of the
total particle number $n_\chi$.  Owing to the relationship between $\chi^2$ and $n_\chi$, our calculations below capture this effect.}

%\begin{figure}[Fig001]
%\centering \leavevmode\epsfysize=2.5cm \epsfbox{\picdir{dotphi.ps }}
%\caption[Fig001]{\label{Fig:dotphi} 
%$|\dot\phi|/(M_pm_{\phi})$ plotted against $m_{\phi}t$ for $g^2=0.1$. The beginning time $t=0$ is chosen at the particle creation moment where $\phi=\phi_0$. The solid red line is output of box simulation. 
%The dashed blue line is solution in a simplified treatment that only consider $\chi$ particle dilution\cite{S}. The dot-dashed black line is inflationary trajectory without particle creation.}
%\end{figure}

\begin{figure}[htbp]
\bigskip \centerline{\epsfxsize=0.25\textwidth\epsfbox{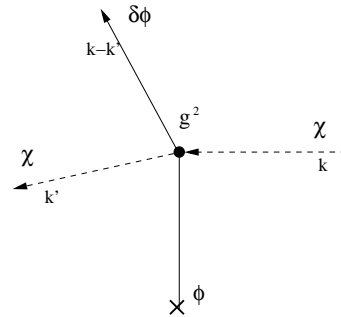}}
%\begin{verse}
%\vskip-0.25cm
\caption{Re-scattering diagram.
}
\label{Fig:diag}
\end{figure}

%\begin{figure}[Fig001]
%\centering \leavevmode\epsfysize=2.5cm \epsfbox{\picdir{diag.eps}}
%\caption[Fig001]{\label{Fig:diag} Re-scattering diagram.}
%\end{figure}

In a parallel development, scalar fields interactions of the type (\ref{coupling}) are the subject of studies in  non-equilibrium QFT and its application to the theory of preheating after inflation, as we mentioned above.
Although we study particle production \emph{during} inflation (as opposed to during preheating, after inflation) there are many similarities.  For example, in the case of parametric resonant preheating due to the oscillating 
inflaton background $\phi(t)$, $\chi$ particles are created in successive bursts whenever $\phi(t)$ crosses zero and the $\chi$ particles become instantaneously massless.  This leads to huge occupation numbers of the created
$\chi$ fluctuations.  On the other hand, in the scenario described above there is only a single burst of particle production and the resulting $\chi$ occupation number (\ref{number}) is always less than unity.
%In this context $\chi$ particles are usually created with the large occupation numbers (for example, 
%in the case of parametric resonant preheating due to the oscillating inflaton background $\phi(t)$, $\chi$ particles are created
%in successive bursts whenever $\phi(t)$ crosses zero). 

The full dynamics of interacting scalars during preheating also includes not only bursts of particle production but also re-scattering effects where $\delta \phi$ fluctuations (particles) are created very quickly due to the 
interaction of created $\chi$ particles with the condensate $\phi(t)$ \cite{KLS97,lattice,FK}.  The diagram for this process is illustrated in Fig.~\ref{Fig:diag}.  The $\delta \phi$ particles produced by re-scattering are far from 
equilibrium and evolve towards an intermediate regime which is well described by the scaling ``turbulent'' solution.  To understand the dynamics, one needs to use lattice numerical 
simulations of time-evolution of the classical scalar fields  based on the LATTICEASY \cite{FT} or DEFROST \cite{Fr} codes, designed for this purpose.
The  turbulent regime of interacting scalars was  investigated in several recent works. The papers \cite{TM,MT} used numerical simulations to demonstrate the scaling regime in the model of
self-interacting classical scalar $\lambda \phi^4$. The papers  \cite{B1,B2} show  numerically the scaling  solution 
for the fully QFT treatment of the same model, and advocate the new regime, the non-thermal fixed point, which may be
asymptotically long (before the system evolves, if ever, to another fixed point: thermal equilibrium).

In this paper we will study in detail the back-reaction of $\chi$ particles produced during inflation on the inflaton field, resulting in Bremsstrahlung radiation and
IR cascading of $\delta \phi$ fluctuations.  Our results will also apply to the early stages of rescattering in preheating after inflation (this is so because the time scale for re-scattering is short 
and hence the results are insensitive to the expansion of the universe).

%\section{Numerical Study of Rescattering}

%{\bf Re-scattering, numerics:}

\section{Numerical Study of Re-scattering}

\begin{figure}[htbp]
\bigskip \centerline{\epsfxsize=0.5\textwidth\epsfbox{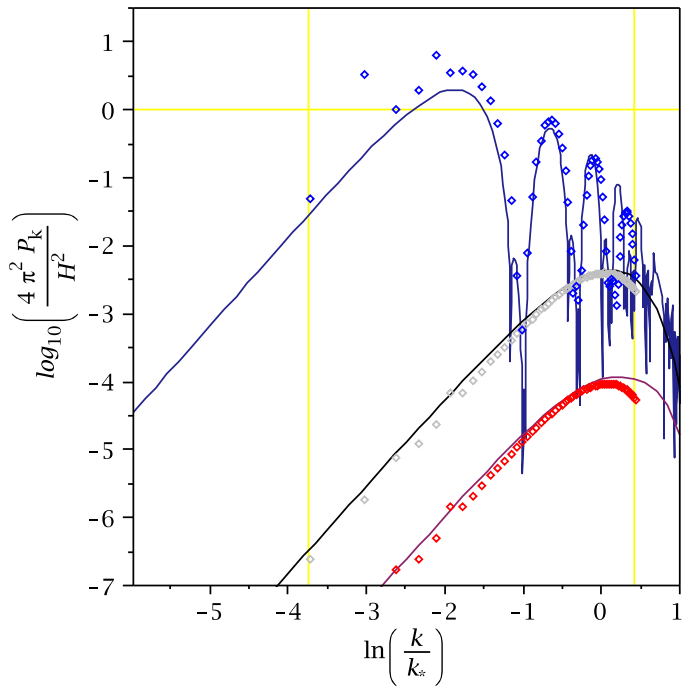}}
%\begin{verse}
%\vskip-0.25cm
\caption{The power spectrum of inflaton modes induced by re-scattering (normalized to the usual vacuum fluctuations) as a function of $\ln(k/k_\star)$, plotted for three representative time steps in the evolution, showing the cascading of power into the IR.
For each time step we plot the analytical result (the solid line) and the data points obtained using lattice field theory simulations (diamonds).  The time steps correspond to the following values of the scale factor: $a = 1.03, 1.04, 2.20$ (where 
$a = 1$ at the moment when $\phi = \phi_0$).  By this time the amplitude of fluctuations is saturated due to the expansion of the universe.
The vertical lines show the range of scales from our lattice simulation.
}
\label{Fig:pwr}
\end{figure}

\begin{figure}[htbp]
\bigskip \centerline{\epsfxsize=0.5\textwidth\epsfbox{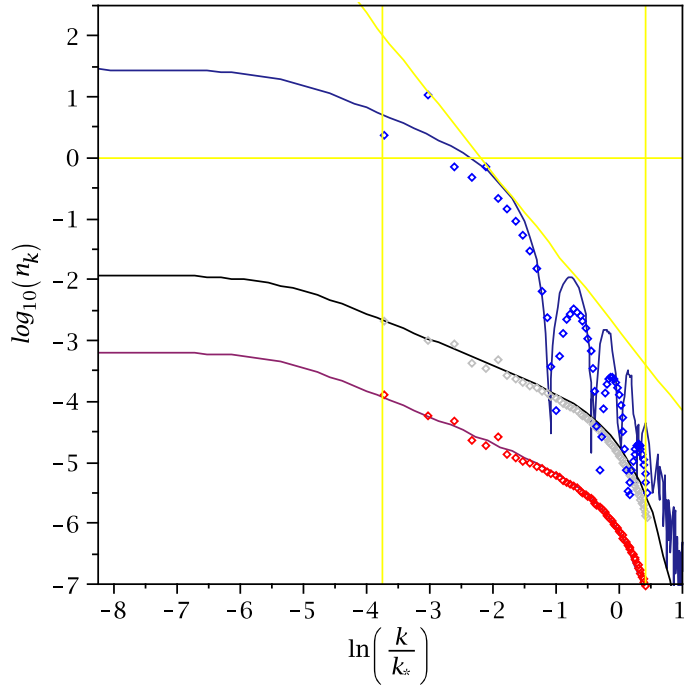}}
%\begin{verse}
%\vskip-0.25cm
\caption{Physical occupation number $n_k$ as a function of $\ln(k / k_\star)$ for $g^2=0.1$.  The three curves correspond to the same series of time steps used in Fig.~\ref{Fig:pwr}, and demonstrate
the growing number of long wavelength inflaton modes which are produced as a result of IR cascading.  Because the same $\chi$-particle can undergo many re-scatterings off the background condensate $\phi(t)$,
the $\delta\phi$ occupation number is larger than the initial $\chi$ particle number (for $g^2=0.1$ one can achieve $n_\phi(k) \sim 30$ even though initially $n_\chi(k) \leq 1$).  When $g^2 = 0.06$ 
the IR $\delta \phi$ occupation number exceeds unity within a single e-folding.
%Note that multiple re-scatterings can produce $\delta\phi$ occupation numbers far exceeding unity, despite the fact that the occupation number of the original $\chi$ particles were always small.
The yellow envelope line shows the early onset of scaling behaviour associated with the scaling turbulent regime.
}
\label{Fig:nk}
\end{figure}

\begin{figure}[htbp]
\bigskip \centerline{\epsfxsize=0.5\textwidth\epsfbox{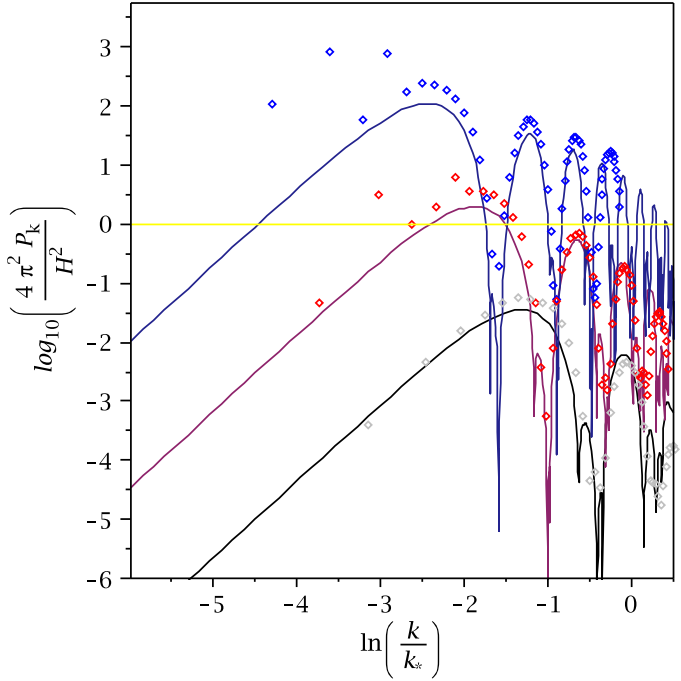}}
%\begin{verse}
%\vskip-0.25cm
\caption{The dependence of the power spectrum $P_\phi$ on the coupling $g^2$.  The three curves correspond to $P_\phi$ for $g^2=0.01,0.1,1$, evaluated at a fixed value of the scale factor, $a = 2.20$. 
We see that even for small values of $g^2$ the inflaton modes induced by re-scattering constitute a significant fraction of the usual vacuum fluctuations after only a single e-folding.
}
\label{Fig:g}
\end{figure}

%In this study we address the question of back-reaction of created particles $\chi$ on the inflaton field resulting in
%the bremschtrahlung radiation  and potential IR cascade  of $\delta \phi$ fluctuations. 
%At first glance, the backreaction in two cases, the instance of particle creation during inflation with the occupation number $n_k \leq 1$
%and explosive   particles creations with $n_k \gg 1$ during preheating, look differently.
To study the creation of $\delta\phi$ fluctuations by re-scattering of produced $\chi$ particles off the condensate $\phi(t)$ in the model (\ref{coupling})
we have adapted the  numerical DEFROST code for the problem of a single burst of instantaneous particle creation during inflation.\footnote{Since the production of long wavelength $\delta \phi$ modes 
is so energetically inexpensive, a major requirement for successfully capturing this effect on the lattice is respecting energy conservation to very high accuracy.  In our modified version of DEFROST 
energy conservation is respected with an accuracy of order $10^{-8}$, compared to $10^{-3}-10^{-5}$ obtained using previous codes.  A minimum accuracy of roughly $10^{-4}$ is required for this problem.}
To run the classical scalar field simulation, 
we must first choose the appropriate initial conditions.  The field $\chi$ on the lattice  is modeled by the random gaussian field realized as the superposition of
planar waves $\chi_k(t) e^{i \vec k \vec x}$ with  random phases.  %At the moment of particle production the temporal eigenfunctions
%are initialized as $\chi_k(0) = 0$, $\dot{\chi}_k(0) = \sqrt{2\omega_k n_k} A_k$ (with $A_k$ are complex gaussian random numbers having statistical magnitude unity).
The initial conditions for the models $\chi_k(t)$ are chosen to emulate the exact quantum mode functions corresponding to the physical occupation number (\ref{number}) (see appendix A for more details) 
while ensuring that the source term for the $\delta \phi$ fluctuations turns on smoothly at $t=0$.
%The temporal eigenfuctions $\chi_k$ are turned on at time $t_0$ from zero to the values $\sqrt{\frac{n_k}{2\omega_k}}$. 
The box size of our $512^3$ simulations corresponds to a comoving scale
which initially is $\frac{20}{2\pi} \sim 3$ times  the horizon size $1/H$, while $k_\star \cong 60 \sqrt{g} H$. 
We run our simulations for roughly 3 e-folding from the initial moment $t_0$ when $\chi$-particles are produced, although a single e-folding
would have been sufficient to capture the effect.  We are  interested in the power spectrum of inflaton fluctuations $P_{\phi}= k^3 \vert \delta \phi \vert^2 / (2\pi^2)$, 
and also the number density of inflaton fluctuations $n_{\phi}(k)= \frac{\Omega_k}{2}\left( \frac{|\delta\dot{\phi}_k|^2}{\Omega_k^2} + |\delta\phi_k|^2 \right)$ (where we introduce 
the notation $\Omega_k = \sqrt{V_{,\phi\phi} + k^2}$ for the inflaton frequency).  For the sake of illustration we have chosen the standard chaotic inflationary potential $V = m^2\phi^2 / 2$
with $m^2 = 10^{-6} M_p$ and $\phi_0 = 3.2 M_p$, however,
our qualitative results will be independent of the choice of background inflation model\footnote{The choice of background inflationary potential will alter the functional form of $\phi(t)$, however, 
all the dynamics of re-scattering occurs within a single e-folding of the moment when $\phi=\phi_0$.  Hence, for any inflation model it will be a good approximation to expand $\phi(t) = \phi_0 + v t$ and our results
should depend only on the velocity $v \equiv \dot{\phi}(t=0)$ (assuming this to be non-zero) which is related to the Hubble scale and the observed amplitude of the curvature perturbation.  The claim of model independence is borne out by explicit analytical
calculations in the next section.  There we find that the dominant contribution to $P_\zeta$ from IR cascading depends sensitively only on the ratio $k_\star / H$ which, as we have shown, is determined entirely by the coupling $g^2$.} and, 
in particular, are applicable to trapped inflation.  We have considered three different values of the coupling constant, 
$g^2 = 0.01, 0.1, 1$, although we focus most of our attention on the case $g^2=0.1$.  %The results of numerical lattice simulations we illustrate in details for the case $g^2=0.1$, while the results for other      $g^2$  ($g^2=1, 0.01$) also will be presented.
Fig.~\ref{Fig:pwr} shows time evolution of the re-scattered inflaton power spectrum $P_{\phi}(k)$ for three different time steps, while Fig.~\ref{Fig:nk} shows the corresponding evolution of
the particle number density $n_{\phi}(k)$.  In Fig.~\ref{Fig:g} we illustrate the dependence of our results on the coupling constant $g^2$.  

%Fig.~\ref{Fig:dotphi} illustrates the impact of rescattering on
%the background velocity $\dot{\phi}$.

In Fig.~\ref{Fig:pwr} we see clearly how multiple re-scatterings lead to a cascading of power into the IR.  These re-scattered inflaton perturbations are complimentary to the usual long-wavelength inflaton modes
produced by quantum fluctuations.  As long as $g^2 > 0.06$ %{\bf * Abstract says $0.06$, what is right?}
the re-scattered power spectrum outside the horizon comes to dominate over the usual vacuum fluctuations within a single e-folding.  At much later times the IR portion
of the power spectrum remains frozen while the UV portion is damped out by the Hubble expansion.  The effect of IR cascading on the power spectrum is much more significant than 
features which are produced by the momentary slowing-down of the background $\phi(t)$.\footnote{To avoid confusion: here we use ``cascading'' to refer to the dynamical process of building up $\delta\phi$ fluctuations in the IR.
If the universe were not expanding, a scaling turbulent regime would be established.  Here we see this scaling regime only in an embryonic form, see the envelope in Fig.~\ref{Fig:nk}.}

Fig.~\ref{Fig:dotphi} illustrates the impact of re-scattering on the dynamics of the velocity of the background field.  The evolution of $\dot{\phi}(t)$ including re-scattering is not changed significantly (as compared to the mean field
theory result), which show the energetic cheapness of IR cascading.

\begin{figure}[htbp]
\bigskip \centerline{\epsfxsize=0.5\textwidth\epsfbox{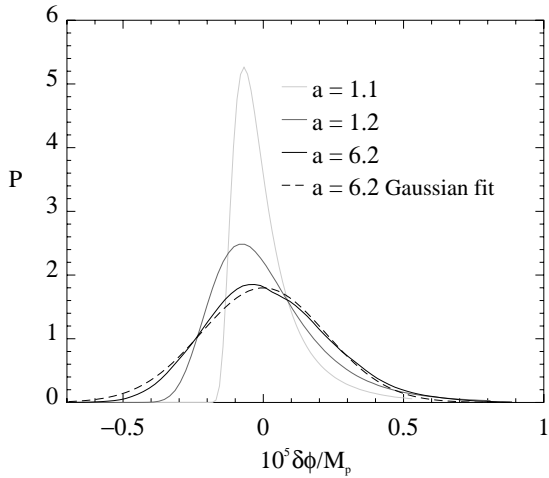}}
%\begin{verse}
%\vskip-0.25cm
\caption{Probability density function of $\delta\phi$ for $g^2=1$ at a series of different values of the scale factor, $a$. The dashed curve shows a Gaussian fit at late time $a=6.2$. 
}
\label{Fig:onepi}
\end{figure}

The long-wavelength inflaton fluctuations produced by IR cascading are non-gaussian.  This is illustrated in Fig.~\ref{Fig:onepi} where we study the probability density function and compare to a gaussian fit.

%{\bf Re-scattering, analytics:}

\section{Analytical Theory of Re-scattering}

We now develop an analytical theory of this effect.  
Here we provide only a cursory discussion, the reader is referred to Appendices A and B for a detailed exposition and technical details of the calculation.
%We neglect the expansion of the universe, split the inflaton field into a classical homogeneous component and quantum inhomogeneities as 
%$\phi(t,{\bf x}) = \phi(t) + \delta\phi(t,{\bf x})$ such that $\langle\phi(t,{\bf x})  \rangle = \phi(t)$ and we further suppose that $\langle \chi(t,{\bf x}) \rangle = 0$.  
At leading order the physics of re-scattering (see Fig.~\ref{Fig:diag})  is described by the equation
\begin{equation}
\label{rescatter_txt}
  \delta\ddot{\phi} + 3H\delta\dot{\phi} - \frac{1}{a^2}\grad^2 \delta\phi + m^2 \delta\phi \cong -g^2 \left[ \phi(t) - \phi_0 \right] \chi^2 \ ,
\end{equation}
where we introduce the notation $m^2 = V_{,\phi\phi}$ for the effective inflaton mass (hence we are not assuming a background potential of the form $m^2 \phi^2 / 2$ in this section, only that $V_{,\phi\phi} \not= 0$
in the vicinity of the point $\phi=\phi_0$).
The solution of (\ref{rescatter_txt}) consists of two components: the solution of the homogeneous equation which simply corresponds to the usual vacuum fluctuations produced during inflation
and the particular solution which is due to the source term.  We will focus our attention on this latter solution which, physically, corresponds to re-scattered inflaton perturbations.
Since the process of IR cascading takes less than a single e-folding, we can safely neglect the expansion of the universe when studying analytically the particular solution of (\ref{rescatter_txt}).
(In all of our lattice simulations the inflationary expansion of the universe is taken into account consistently.)
Solving for the particular solution $\delta\phi_k$ of (\ref{rescatter_txt}) and defining the re-scattered power spectrum $P_\phi$ in terms of the QFT correlation function in the usual manner we arrive 
at an expression for $P_\phi$ in terms of the c-number mode functions $\chi_k$ which obey equation (\ref{field1}).\footnote{We are only interested in connected contributions to the correlation functions,
which is equivalent to subtracting the expectation value from the source term in (\ref{rescatter_txt}): $\chi^2 \rightarrow \chi^2 - \langle\chi^2\rangle$.  Thus, our re-scattered inflaton modes are only
sourced by the variation of $\chi^2$ from the mean $\langle\chi^2\rangle$.}  The result is
\begin{eqnarray}
  P_\phi  &=& \frac{g^4 \dot{\phi}_0^2}{8\pi^5}\frac{k^3}{\Omega_{k}^2} \int dt'dt''t't''\sin\left[\Omega_{k}(t-t')\right]\sin\left[\Omega_{k}(t-t'')\right] \nonumber \\
 && \,\,\,\,\,\,\,\,\,\, \times \int d^3k' \chi_{k-k'}(t')\chi_{k-k'}^\star(t'') \chi_{k'}(t')\chi_{k'}^\star(t'') \label{pwr_phi_txt} %{pwr_phi}
\end{eqnarray}
where again we have $\Omega_k  = \sqrt{k^2 + m^2}$ for the $\delta\phi$-particle frequency.

To evaluate this power spectrum we need an expression for the solutions of (\ref{field1}) in the regime of interest.  Let us choose the origin of time so that $t=0$ corresponds to the moment when $\phi = \phi_0$.
At the moment $t=0$ the parameter $|\dot{\omega}_k| / \omega_k^2$ is order unity or larger and $\omega_k$ varies non-adiabatically.  At this point $\chi_k$ modes are produced in the momentum band $k \lsim k_\star$.  
However, within a time $\Delta t \sim k_\star^{-1}$ (which is tiny compared to the Hubble time $H^{-1}$) the $\chi$ particles become extremely heavy and their frequency again varies adiabatically.  At times $t \gsim k_\star^{-1}$ 
we can safely approximate $\omega_k = \sqrt{k^2 + k_\star^4 t^2} \cong k_\star^2 t$ for the modes of interest and $\chi_k$ takes the simple form
\begin{equation}
\label{chi_out_txt}
  \chi_k(t) \cong \sqrt{1 + n_k} \,\, \frac{e^{-i(k_\star t)^2 / 2}}{k_\star \sqrt{2t}} -i \sqrt{n_k} \,\, \frac{e^{+i(k_\star t)^2 / 2}}{k_\star \sqrt{2t}} \ ,
\end{equation}
where the occupation number was defined in (\ref{number}).  The factors $\sqrt{1 + n_k}$, $-i \sqrt{n_k}$ are the Bogoliubov coefficients while the factors proportional to $e^{\pm i(k_\star t)^2 / 2}$ come from the positive and 
negative frequency adiabatic mode solutions \cite{KLS97}.  As we see, very quickly after $t=0$ the $\chi$ particles become very massive and their multiple re-scatterings off the condensate $\phi(t)$ generates Bremsstrahlung radiation of
IR $\delta \phi$ particles.

We have computed the full renormalized power spectrum analytically in closed form and the result is presented in equation (\ref{full_pwr_result}).  This formula is used for all of our figures.
Since the exact analytical result is quite cumbersome, it is useful to consider the following representative contribution to (\ref{pwr_phi_txt}):
\begin{equation}
\label{pwr_rep}
  P_\phi \simeq \frac{g^2\, k^3 k_\star^3}{32\sqrt{2}\, \pi^5} \left[\frac{1-\cos(\Omega_k t)}{\Omega_k^2}\right]^2 e^{-\pi k^2 / (2k_\star^2)} \ , % \exp\left[-\frac{\pi k^2}{2 k_\star^2}\right]
\end{equation}
which captures the properties of the full analytical solution.  In particular, the simple expression (\ref{pwr_rep}) nicely describes the IR cascade. The spectrum has a peak which initially (near $t\sim k_\star^{-1}$) 
is close to $k_\star$.  As time evolves the peak moves to smaller-and-smaller $k$ as power builds up in the IR.  From (\ref{pwr_rep}) we see that modes with $\Omega_k t  < 1$ gain power as $P_\phi(k) \sim t^4$.  For a given
$k$-mode the growth of the power spectrum saturates when $\Omega_k t\sim 1$, however, the cascade still continues at some lower $k$.  If we had $m=0$ then the cascade would continue forever, otherwise 
formula (\ref{pwr_rep}) predicts that the growth of the spectrum saturates at $t \sim m^{-1}$ when the peak has reached $k \sim m$.  After this point the character of the IR cascade is expected to change, however, our analytic 
calculation is no longer reliable because $m \ll H$ and we have neglected the expansion of the universe.  Notice from equation (\ref{pwr_rep}) that the value of $P_\phi / H^2$ at the peak (which is the main observable signal) is fixed by the ratio
$k_\star / H$, which in turn depends only on the coupling $g^2$ (there is some dependence also on $m^2 = V_{,\phi\phi}$, however, this is subdominant since $m^2 \ll H^2$ for any inflation model).  This observation confirms our previous claim that the
dynamics of IR cascading are largely insensitive to the choice of background inflation model.

\begin{figure}[h!]
\begin{center}
\subfigure%[\, Single burst.\label{sub1}]
{\includegraphics[scale=0.5]{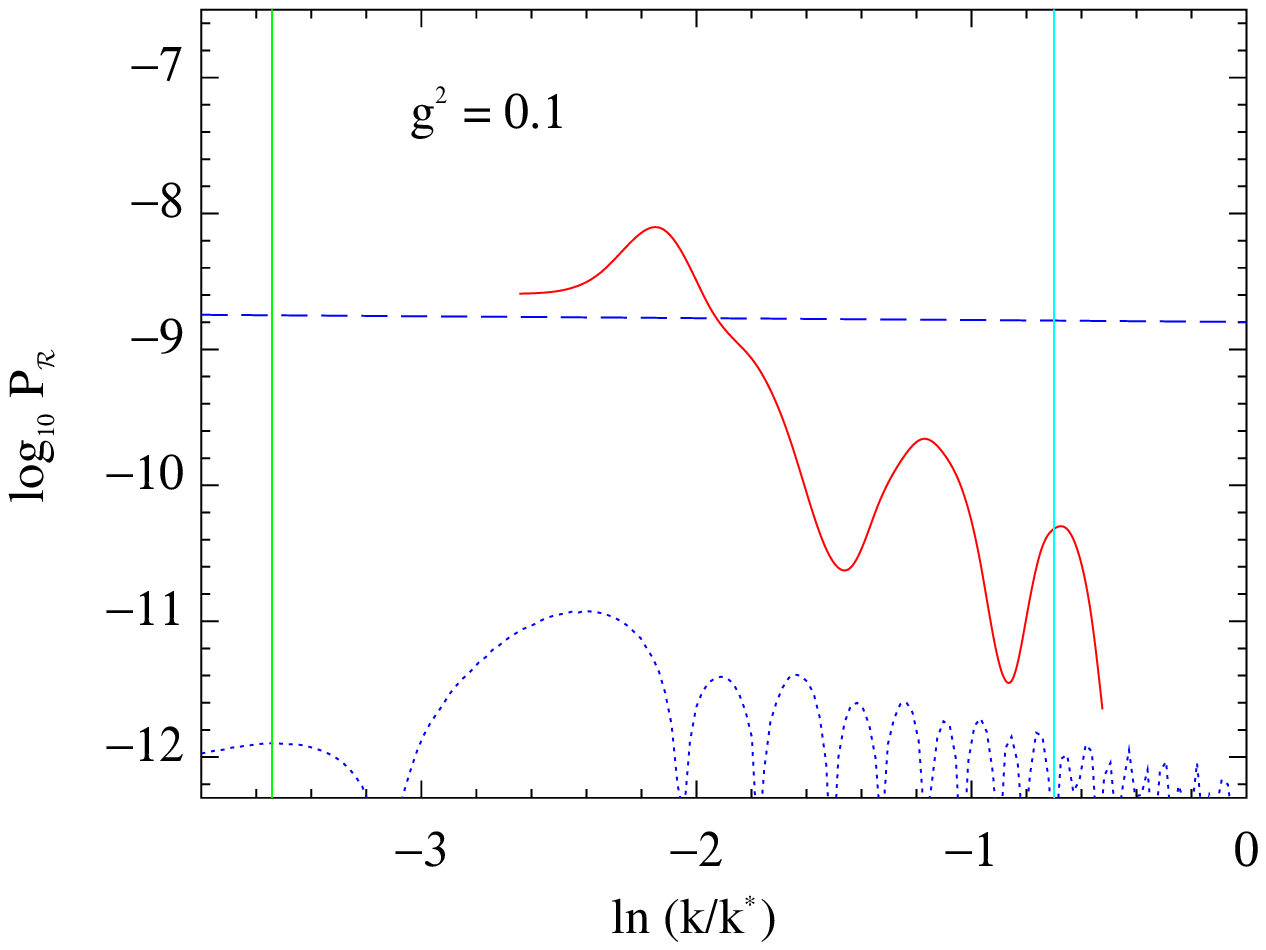}}
\subfigure%[\, Multiple bursts.\label{sub2}]
{\includegraphics[scale=0.5]{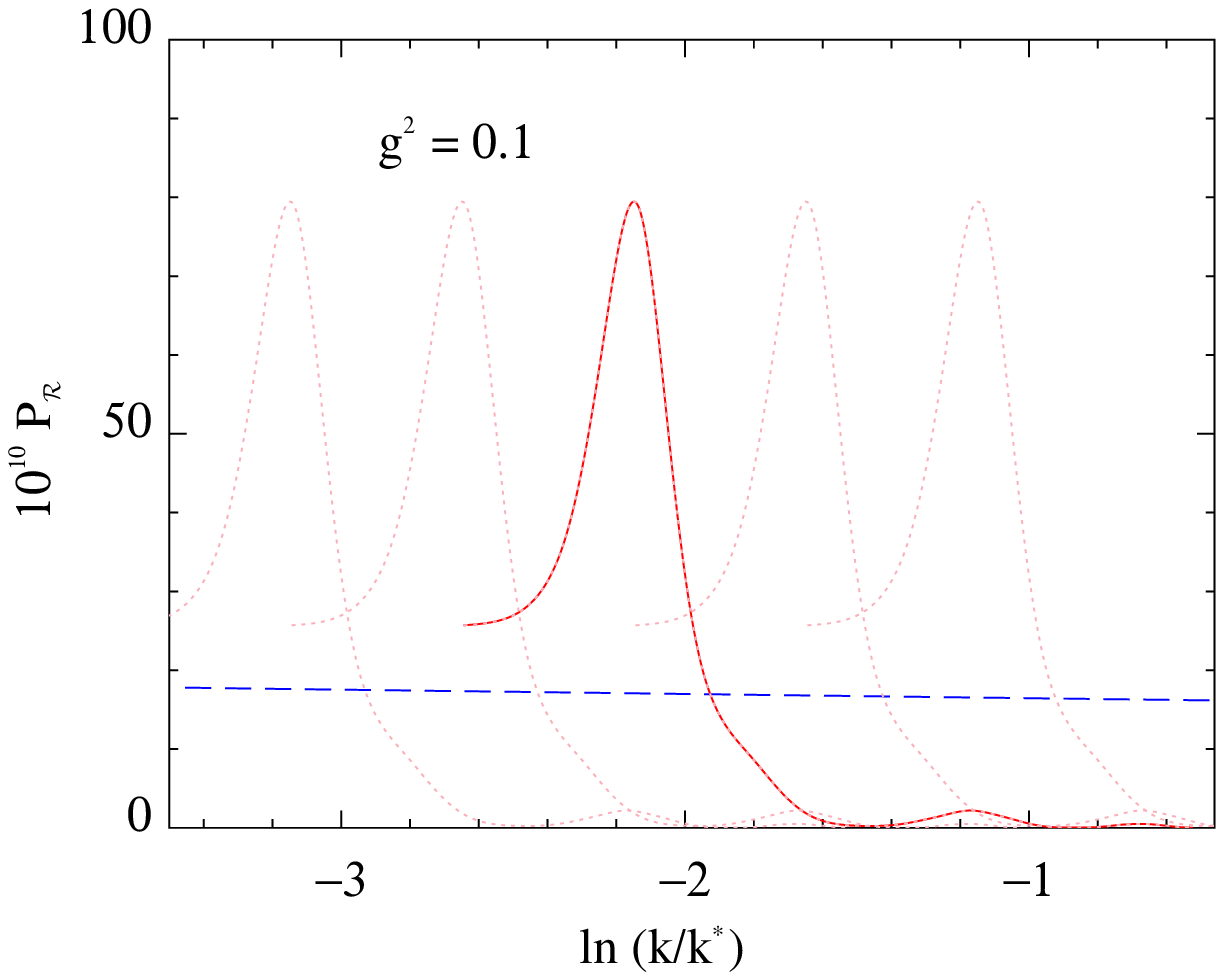}}
\caption{\label{Fig:curvature} 
The top panel shows a comparison of curvature fluctuations from different effects.  We see the dominance of fluctuations produced by IR cascading over the wiggles induced by the momentary slowing-down of the inflaton.
For illustration we have taken $g^2=0.1$, but the dominance is generic for all values of the coupling.  The red solid line is the IR cascading curvature power spectrum, while the blue dashed line is the result of a mean field treatment.
(The vertical lines show $aH$ at the beginning of particle production and after $\sim 3$ e-foldings.)
The bottom panel shows the curvature power spectrum resulting from multiple bursts of particle production and IR cascading.  Superposing a large number of these bumps produces a broad-band spectrum.
%The top panel shows the dominance of fluctuations produced by IR cascading over the wiggles induced by the momentary slowing-down of the inflaton.  The red solid line is the curvature power spectrum due to rescattering
%while the blue dashed line is the result of a mean field treatment which neglects rescattering.  This latter curve contains the wiggles that were studied in \cite{S}, however, they are not visible on this plot because their smallness.
%The black dotted line is the difference between the rescattering spectrum and the mean field result, showing the dominance of the former.  The vertical lines show $aH$ at the beginning of particle production and $\sim 3$ e-foldings
%later.  The bottom panel displays the curvature power spectrum resulting from multiple bursts of particle production and IR cascading.  Superposing a large number of these bumps produces a smooth, scale-invariant spectrum.
}
\end{center}
\end{figure}

%{\bf Discussion of  Curvature Fluctuations from IR Cascading:} 

\section{Discussion of  Curvature Fluctuations from IR Cascading}

Any inflaton fluctuations $\delta \phi$, independently on their origin,
evolve qualitatively similarly during inflation. When  their physical 
wavelength is smaller than the Hubble radius $1/H$, $\delta \phi$ is oscillating 
while their amplitude is diluted as $1/a$. As far as the wavelength exceeds the Hubble radius,
the amplitude of $\delta \phi$ freezes out. Fluctuations of $\delta \phi$ induce the curvature metric fluctuations.
Inflationary expansion of the universe further stretch the wavelengths of the fluctuations frozen outside the horizon,
making them potentially of the cosmological scales, depending on the wavelength.
The inflaton fluctuations produced by the IR cascading, therefore, are the 
potential sources for observable curvature fluctuations.
To calculate curvature fluctuations generated by the IR cascading, we have 
to solve self-consistent system of linearized
Einstein equations for metric and the fields fluctuations.
%For instance, let us  chose to work in the longitudinal gauge where the 
%background  geometry perturbed by the scalar
%metric  fluctuations $\Phi$ has the form 
%\begin{equation}\label{scalar}
%ds^2=a^2 \left[ (1+2\Phi)d\tau^2-(1-2\Phi) d{\vec x^2}\right] \ ,
%\end{equation}
%where $\tau$ is the conformal time, and below $'$ will be derivative with 
%its respect.
For example, the $(0,0)$ linearized Einstein equation for our model  reads 
%as {\bf check! I don't have books at home}
\begin{equation}\label{einsten}
  \delta G_0^0 =\frac{8\pi}{M_p^2}\left(\delta 
T^0_0(\phi)+\delta T^0_0(\chi) \right) \ .
\end{equation}
where $\delta G_0^0$ is the perturbed Einstein tensor and in the {\it r.h.s.}~$\delta T^0_0(\phi)$ corresponds to the fluctuations of 
the inflaton energy density, containing familiar terms linear with respect to $\delta \phi$, like $\dot \phi \delta \dot \phi$ etc. Second term corresponds to the 
contribution from $\chi$ particles
\begin{equation}\label{second}
\delta T^0_0(\chi)=\frac{1}{2}\dot \chi^2 +\frac{1}{2} (\nabla 
\chi)^2+\frac{1}{2}g^2 (\phi-\phi_0)^2 \chi^2-<T^0_0(\chi)> \ .
\end{equation}
Although this expression is bi-linear {\it w.r.t.}~$\chi$, it turns out 
taking $\delta T^0_0(\chi)$ into account is important.
To begin  the investigation  the equation (\ref{einsten}), it is convenient 
to use its Fourier transformation.
While the Fourier components of $\delta T^0_0(\phi)$ contains linear terms 
of $\delta \phi_k$, the Fourier transform
of $\delta T^0_0(\chi)$ contains convolutions like $\sim \int d^3 {\vec k'} 
\dot \chi_{\vec k'} \dot \chi^*_{\vec k-\vec k'}$. etc.
As a result, despite the fact that $\chi$ particles amplitude is peaked at 
$k \sim k_\star$, this type of convolution gives
significant contribution at small $k$, which are of interest for the theory 
of generation of cosmological fluctuations.
Preliminary estimations based on the analytical formulas for $\chi_k$ 
involved in the convolution, show that
contribution of $\delta T^0_0(\chi)$ is at least the same order of
magnitude as $\delta T^0_0(\phi)$.
Rigorous treatment of the curvature fluctuations in our model is therefore 
rather complicated, and will be leave it for separated project.

Since both terms in {\it r.h.s.}~in (\ref{einsten}) are  the same order  of 
magnitude, here for the crude estimations we will use the simple-minded formula $P_{\zeta} \sim \left(\frac{H^2}{\dot \phi} \right)^2$.
The curvature fluctuations generated by the IR cascading, are illustrated in the top panel of Fig.~\ref{Fig:curvature}.
The curvature fluctuations from the instance of the IR cascading has the 
bump-like shape within the interval
of the wavelength, roughly corresponding to one e-folding.
They are significantly, by orders of magnitude,
dominated
over the fluctuations generated by the momentary slowing down of $\phi(t)$.
If we pick up the background inflationary model to be chaotic inflation with 
the standard quadratic potential,
the ratio of the power spectra from IR cascading and the standard 
fluctuations is estimates as $P_{IR}/P_s \sim 700 \times g^{4.5}$.  Thus, depending on the coupling $g^2$, 
the IR bump can dominate (for $g^2 > 0.06$)
over the standard fluctuations, or just contribute to them for smaller 
$g^2$.

Suppose that we have a sequence of the particles creation events at different 
moments $t_{0i}$, $i=1,2,3,...$.
Each of those events generate, through IR cascading, corresponding bumps in 
the spectrum, as illustrated in the lower panel of Fig.~\ref{Fig:curvature}.
Depending on the density of $t_{0i}$ moments, superposition of such IR bumps 
results in broad band contribution to
the curvature power spectrum.

We also estimated non-gaussianity of $\delta \phi$ fluctuations from IR 
cascading. They are quite significant, we estimate
the non-gaussianity parameter  $f_{NL} \sim 2\times 10^4  \, g^{2.25}$.
Therefore the non-gaussian signal from individual bump can be strongly 
non-gaussian.
In the model with multiple instances of particle creations,
the broad-band IR cascading fluctuations dominated over the standard
fluctuations, apparently, are ruled because of the strong non-gaussianity. 
However, the broad-band IR cascading fluctuations
can be considered as additional subdominant component to the standard 
fluctuations. In this case the
non-gaussianity of the net curvature fluctuations can be acceptable but 
different from that of the standard fluctuations alone.  We leave a detailed discussion
of the nongaussianities produced during IR cascading (and their observability) to future studies.
Notice that this type of nongaussianity, which is localized over a narrow region of scales, is similar to what
was obtained in \cite{spikes}.

Another important parameter of the IR cascading fluctuations is the 
wavelength of the bump, which depend on the value of
$\phi_0$. there are interesting possibilities to consider them at small CMB 
angular scales (small-scale non-gaussianity?), at scales of galaxies, at near the horizon scales (CMB anomalies at large 
scales?).
We leave all of these possibilities for future discussion.

%{\bf Summary:}  

\section{Summary and Conclusions}

We find the following new results for interacting scalars during inflation in the model (\ref{coupling}).\\
i) In the early stages of re-scattering, when the back-reaction can be treated linearly,
the spectrum of inflaton fluctuations $\delta \phi$  and the corresponding particle number density $n_{\phi}(k)$ can 
be rigorously calculated with QFT with the diagram in Fig.~\ref{Fig:diag}. We perform such QFT calculations and compare 
with lattice simulations of the classical field dynamics.  The results are highly compatible with each other,
even well into the late time nonlinear regime.
%Indeed, extrapolation of the analytic result based on the Fig.~\ref{Fig:diag} even into the non-linear late time regime
%still predicts a result which is compatible with that of full lattice simulation. 
This signals the dominance of Fig.~\ref{Fig:diag} in the dynamics of re-scattering, while the analytic estimate gives a handy fitting formula.\\
ii) While the stationary scaling turbulent solution for the scalar fields after preheating was established previously, the way this regime
appears dynamically was not traced out in detail.  For our example 
for the first time we show explicitly how this scaling regime is seeded.
%we show how the scaling regime emerges, what is the timing of IR
%cascade propagation and what is the profile of the spectra resulting from the IR cascading.  
In the absence of expansion of the universe, this embryonic
scaling behaviour will develop into the full turbulent regime that has previously been observed, however, for our purposes only the early stages are relevant because fluctuations
freeze outside of the horizon. \\
iii) The most unexpected result, which is of interest outside of the inflationary theory, is that 
even an insignificant amount of out-of-equilibrium particles with $n_{\chi}(k) \leq 1$, being 
re-scattered off the scalar field condensate,  can generate IR cascade  of the inhomogeneous condensate fluctuations 
with large occupation number $n_{\phi}(k)$ in the IR region. This is explained by fact that multiple production of the IR modes is energetically cheap.\\
iv) IR fluctuations of the light fields have special significance in the context of inflationary theory.
These fluctuations evolve in time similar to the evolution of the usual inflationary fluctuations. Their
amplitude is oscillating while their  wavelengths is inside the Hubble radius and is frozen out once their wavelengths exceed the Hubble radius $H^{-1}$. 
However, the amplitude of IR cascading fluctuations is different from that of the usual quantum fluctuations.  Frozen fluctuations $\delta \phi$,
regardless of their origin, %(either from the initial vacuum fluctuations or from IR cascading), 
will induce cosmological curvature fluctuations.  Thus,  we get  a new mechanism for generating frozen long wavelength $\delta \phi$ fluctuations from IR cascading.
Therefore, IR cascading will lead to observable features in the CMB power spectrum.  For generic choices of parameters, these re-scattered 
fluctuations are much more significant than the features induced by the momentary slowing-down of the background $\phi(t)$, see the upper panel of Fig.~\ref{Fig:curvature}.\\
v) Since the solution $\delta\phi$ of (\ref{rescatter_txt}) depends nonlinearly on the gaussian field $\chi$, the curvature fluctuations induced by IR cascading
will be non-gaussian.  This nongaussianity is illustrated in Fig.~\ref{Fig:onepi}.  We estimate this nongaussianity to be significant.
%For the model of quadratic chaotic inflation preliminary estimations give 
%an effective non-gaussian parameter $f_{NL} \sim 2 \times 10^4 g^{2.25}$, which is pretty big.  
However, this non-gaussian signal is related to the IR cascading bump of the spectrum
and peaks on the range of scales corresponding to roughly one e-folding after $t=t_0$.  This type of nongaussianity, which is large only over a small range of scales, is not well constrained by observation.\\
vi) The strength and location of our effect is model-dependent (through $g^2$ and $\phi_0$), however, the very fact that subtle QFT effects of interaction during inflation may lead to an
observable effect is intriguing.\\
vii) In our analysis we have focused on a single burst of instantaneous particle production during inflation.  This scenario is interesting in its own right, however, our results could also 
be extended in a straightforward manner to study trapped inflation models where there are numerous bursts of particle production; see \cite{E} for more detailed discussion.\\
viii) Suppose we have a sequence of points $\phi_{0i}$ ($i=1,\cdots,N$) where particles $\chi_i$ become massless.  In this case the curvature fluctuation profiles generated from individual bursts
of IR cascading can superpose to form a smooth spectrum of cosmological fluctuations, see the lower panel of Fig.~\ref{Fig:curvature}.  This provides us with a new mechanism for generating long wavelength curvature fluctuations during inflation
from IR cascading.  The amplitude and non-gaussianity of these curvature fluctuations will depend on the coupling, $g^2$. 
These fluctuations are interesting on their own, although they may generate too much non-gaussianity. They also can be considered as an extra component of the standard vacuum fluctuations, introducing an interesting non-gaussian signal to the
net fluctuations.\\
ix) The transfer of energy into fluctuations via successive bursts of particle production can lead to trapped inflation.   Our new mechanism of generating cosmological fluctuations from IR cascading can, but need not, be
 associated with trapped inflation.\\
x) Varying the location, strength and non-gaussianity of the IR cascading bump, it will be interesting to consider 
other potential implication to the cosmological fluctuations, e.g.\ their impact on the
horizon scale fluctuations or on small scale fluctuations where they might effect primordial Black Hole formation or the
generation of gravitational waves.

Finally, let us return to the old story of how cosmological fluctuations are affected by phase transition during inflation, which we discussed at the beginning of this paper.
We project that our results concerning re-scattering and IR cascading will radically change the conventional picture.  %We plan to consider this in a forthcoming publication.

We thank Juergen Berges, Dick Bond, Andrei Frolov, Andrei Linde, Antonio Enea Romano, Misao Sasaki, David Seery and Eva Silverstein for useful discussions.  N.B.,  L.K.\ and D.P.\ were supported by NSERC; L.K.\ was also supported by CIFAR.
D.P.\ thanks CITA for  hospitality under CITA Senior Visitors Program.

\renewcommand{\theequation}{A-\arabic{equation}}
\setcounter{equation}{0}
\section*{APPENDIX A: Analytical Theory of Re-scattering}
\label{appA}

\renewcommand{\thesubsection}{A.\arabic{subsection}}
\setcounter{subsection}{0}

In this appendix we develop an analytical theory of re-scattering which is in good agreement with the result of fully nonlinear lattice field theory simulations.
As usual we split the inflaton field into a classical homogeneous component and quantum inhomogeneities as $\phi(t,{\bf x}) = \phi(t) + \delta\phi(t,{\bf x})$ such that 
$\langle\phi(t,{\bf x})  \rangle = \phi(t)$ and we further suppose that $\langle \chi(t,{\bf x}) \rangle = 0$.  Since IR cascading occurs within a single e-folding we
can safely neglect the expansion of the universe.  However, there is no obstruction to consistently including this effect \cite{neilprog}.

At leading order the physics of re-scattering is described by equation (\ref{rescatter_txt}), 
%\begin{equation}
%\label{rescatter}
%  \delta\ddot{\phi} - \grad^2 \delta\phi + m^2 \delta\phi \cong -g^2 \left[ \phi(t) - \phi_0 \right] \chi^2 \ .
%\end{equation}
which corresponds to the diagram in Fig.~\ref{Fig:diag}.  There is a correction to (\ref{rescatter_txt}) corresponding to a diagram where two $\delta\phi$ particles interact with two $\chi$
particles, however, this effect is sub-leading \cite{KLS97}.  It is understood that one must subtract from (\ref{rescatter_txt}) the expectation value of the right-hand-side in order to consistently define the 
quantum operators $\delta\phi$ such that $\langle\delta\phi\rangle = \langle\chi\rangle = 0$.  Subtracting off this expectation value is equivalent to only considering connected diagrams 
when we compute correlation functions.

\subsection{Production of $\chi$-Particles}

To solve equation (\ref{rescatter_txt}) we first require explicit expressions for the background field $\phi(t)$ and the wavefunction $\chi(t,{\bf x})$.  Let us choose the origin of time so that
$\phi = \phi_0$ at $t=0$.  Near the moment of particle production  we can expand $\phi(t) - \phi_0 \cong \dot{\phi}_0 t$.
The interaction term in (\ref{coupling}) induces induces a mass for the $\chi$-field
\begin{equation}
\label{m_chi}
  m_\chi^2 = g^2 \left[\phi(t)-\phi_0 \right]^2 \cong g^2 \dot{\phi}^2 t^2 \equiv k_\star^4 t^2
\end{equation}
which vanishes at $t=0$.  At this moment particles will be copiously produced by quantum effects.

The mode functions $\chi_k(t)$ obey the following equation
\begin{equation}
\label{chi_prod}
  \ddot{\chi}_k(t) + \omega_k^2(t) \chi_k(t) = 0
\end{equation}
where the time-dependent frequency is
\begin{equation}
\label{omega_chi}
  \omega_k(t) = \sqrt{ k^2 + m_\chi^2  } = \sqrt{ k^2 + k_\star^2(k_\star t)^2  }
\end{equation}
The theory of equation (\ref{chi_prod}) is well-studied in the literature \cite{KLS97,B}.  As long as the frequency (\ref{omega_chi}) varies adiabatically $|\dot{\omega}_k| / \omega_k^2 \ll 1$
the modes of $\chi$ will not be excited and are well described by the adiabatic solution $\chi_k(t) = f_k(t)$ where we have defined
\begin{equation}
\label{f_k}
  f_k(t) \equiv \frac{1}{\sqrt{2\omega_k(t)}} \exp\left[-i \int^t dt' \omega_k(t') \right]
\end{equation}
However, very close to $t=0$, roughly within the interval $-k_\star^{-1} < t < +k_\star^{-1}$, the parameter $|\dot{\omega}_k| / \omega_k^2$ can become order unity or larger for low momenta 
$k \lsim k_\star$ and $\chi_k$ modes within this band will be produced.  The general solution of (\ref{chi_prod}) can be written in terms of the adiabatic modes (\ref{f_k}) and the time-dependent 
Bogoliubov coefficients as
\begin{equation}
\label{chi_bog}
  \chi_k(t) = \alpha_k(t) f_k(t) + \beta_k(t) f_k(t)^\star
\end{equation}
where the Bogoliubov coefficients obey a set of coupled ordinary differential equations with initial conditions $|\alpha_k(0^{-})| = 1$, $\beta_k(0^{-}) = 0$.   Near $t=0$ the adiabaticity condition 
is violated and $\beta_k$ grows rapidly away from zero as a step-like function.  Very shortly after this burst of particle production the frequency again varies adiabatically and $\alpha_k$, $\beta_k$ 
become constant, taking the following values \cite{KLS97}:
\begin{eqnarray}
  \alpha_k(t>0) &=& \sqrt{1 + n_k} \label{alpha} \\
  \beta_k(t>0) &=& \sqrt{ n_k  }\,\, e^{i\delta_k} \label{beta}
\end{eqnarray}
where the physical occupation number is defined by (\ref{number}).  The phase $\delta_k$ has been computed analytically in \cite{KLS} and depends nontrivially on $k$.  However, since most of the particle 
production occurs for momenta $k \lsim k_\star$ it is an excellent approximation to use the simple result $e^{i \delta_k} \cong -i$. (We have verified that changing the relative phase will at most alter factors 
order unity in the final results.)

%The phases $\delta_k$, $\theta_k$ play an important role in the case where $\phi(t)$ passes through zero multiple times, as in the theory 
%of preheating after inflation.  For the case we consider, a single burst or particle production, we can simply neglect the phases.

We are now in a position to write out the solution for the $\chi_k$ modes in the outgoing adiabatic regime $t \gsim k_\star^{-1}$.  Since most of our interest is in IR modes with $k \lsim k_\star$ it
is a good approximation to expand the frequency (\ref{omega_chi}) as $\omega_k(t) \cong k_\star(k_\star t)$.  Using the equations (\ref{alpha}) and (\ref{beta}) we can write the solution (\ref{chi_bog}) in 
the region of interest as
\begin{equation}
\label{chi_out}
  \chi_k(t) \cong \sqrt{1 + n_k} \,\, \frac{e^{-i(k_\star t)^2 / 2}}{k_\star \sqrt{2t}} - i \sqrt{n_k} \,\, \frac{e^{+i(k_\star t)^2 / 2}}{k_\star \sqrt{2t}}
\end{equation}

\subsection{Equations for Re-scattering}

Having reviewed the solutions for $\phi(t)$ and $\chi_k(t)$ we now turn our attention to solving (\ref{rescatter_txt}).
Let us first briefly discuss our conventions for fourier transforms and mode functions.  We write the q-number valued 
fourier transform of $\chi$ as
\begin{equation}
\label{chi_fourier}
  \chi(t,{\bf x}) = \int \frac{d^3k'}{(2\pi)^{3/2}} e^{i{\bf k}\cdot {\bf x}} \xi^\chi_{\bf k}(t)
\end{equation}
Because $\chi$ is gaussian we can expand $\xi^\chi_{\bf k}$ into c-number mode functions $\chi_k$ (discussed above) and annihilation/creation operators $a_{\bf k}$, $a^\dagger_{\bf k}$ as
\begin{equation}
\label{chi_mode}
  \xi^\chi_{\bf k}(t) = a_{\bf k}\, \chi_{ k}(t) + a_{-{\bf k}}^\dagger\, \chi_{{ k}}^\star(t)
\end{equation}
In the theory of preheating/moduli trapping without re-scattering the distinction between q-number fourier transform and c-number mode functions is not important because both obey the same 
equation of motion (equation (\ref{chi_prod}) in the case at hand).  However, once re-scattering is taken into account this distinction is crucial.  To see why, note that the solution $\delta\phi$ 
of equation (\ref{rescatter_txt}) will not be gaussian and hence will not admit an expansion of the form (\ref{chi_mode}).

Finally, we return to the equation for re-scattering, eqn.\ (\ref{rescatter_txt}).  We can solve for the q-number fourier transform of $\delta \phi$ (defined analogously to (\ref{chi_fourier})) using the retarded Green
function
\begin{eqnarray}
  \xi^\phi_{\bf k}(t) &=& \frac{g^2 \dot{\phi}}{(2\pi)^{3/2}}\frac{1}{\Omega_k}\int_0^tdt't'\sin\left[\Omega_k(t-t')\right]\nonumber \\
  && \,\,\,\,\,\,\,\,\times \int d^3k' \xi^\chi_{{\bf k}-{\bf k'}}(t')\xi^\chi_{\bf k'}(t') \label{gf_soln}
\end{eqnarray}
where we have introduced the notations $\Omega_k  = \sqrt{k^2 + m^2}$ for the $\delta\phi$-particle frequency and $m^2 = V_{,\phi\phi}$ for the effective $\delta\phi$ mass.  
Carefully carrying out the Wick contractions yields
\begin{eqnarray}
  \langle \xi^\phi_{k_1}(t)\xi^\phi_{k_2}(t) \rangle &=& \frac{2 g^4 \dot{\phi}^2}{(2\pi)^3}\frac{1}{\Omega_{k_1}^2} \delta^{(3)}({\bf k_1}+{\bf k_2}) \nonumber \\
                                                     &&\hspace{-22mm}\times \int dt'dt''t't''\sin\left[\Omega_{k_1}(t-t')\right]\sin\left[\Omega_{k_1}(t-t'')\right] \nonumber \\
                                                     &&\hspace{-22mm} \times \int d^3k' \chi_{k_1-k'}(t')\chi_{k_1-k'}^\star(t'') \chi_{k'}(t')\chi_{k'}^\star(t'') \label{pwr_phi}
\end{eqnarray}
where the $\chi$-particle mode functions $\chi_k$ are defined by (\ref{chi_mode}).  Defining the power spectrum in terms of the two-point function in the usual manner 
\begin{equation}
    \langle 0 | \xi^\phi_{\bf k}(t) \xi^\phi_{\bf k'}(t) | 0 \rangle \equiv \delta^{(3)}( {\bf k} + {\bf k'} ) \frac{2\pi^2}{k^3} P_\phi \label{corr2}
\end{equation}
we can extract the power in re-scattered $\phi$ modes.

Alternatively one could compute the power spectrum of re-scattered inflaton modes using the Schwinger's ``in-in'' formalism \cite{in-in} which was implemented to compute cosmological perturbations by Weinberg in \cite{weinberg}.
We have verified that the tree level contribution to $P_\phi$ obtained using this formalism reproduces our result (\ref{pwr_phi}).  Our approach is analogous to computing the cosmological perturbation from the field equations
using the Seery et al.\ approach \cite{seery}.  The consistency of this method with the in-in approach at tree level is in accordance with the general theorem of \cite{tree}.

\subsection{Renormalization}

To compute the spectrum of re-scattered $\delta \phi$-particles we simply need to insert the solution (\ref{chi_out}) into (\ref{pwr_phi}) and evaluate the integrals.   However, there is one subtlety.  The resulting
power spectrum is formally infinite, moreover, it contains the effect of both particle production as well as vacuum fluctuations of the $\chi$ field.  We are only interested in the re-scattered $\delta\phi$ which are 
due to particle production, thus, we need to subtract off the contribution due to nonlinear $\delta\phi$ production by $\chi$ vacuum fluctuations.  

To properly define the two-point function of $\delta\phi$ we need to renormalize the four-point function of the gaussian field $\chi$.  As a warm-up, let us first consider how to renormalize the two point function
of the gaussian field $\chi$.  We use the following scheme
\begin{equation}
  \langle \xi^\chi_{k_1}(t_1) \xi^\chi_{k_2}(t_2) \rangle_{\mathrm{ren}} = \langle \xi^\chi_{k_1}(t_1) \xi^\chi_{k_2}(t_2) \rangle - \langle \xi^\chi_{k_1}(t_1) \xi^\chi_{k_2}(t_2) \rangle_{\mathrm{in}}
\end{equation}
where $\langle \xi^\chi_{k_1}(t_1) \xi^\chi_{k_2}(t_2) \rangle_{\mathrm{in}}$ is the contribution in the absence of particle production, computed by simply taking the solution (\ref{chi_bog}) with $\alpha_k=1$, $\beta_k=0$.
More explicitly, for the case at hand, we have
\begin{eqnarray}
  \langle\chi^2(t,{\bf x})\rangle_{\mathrm{ren}} &=& \int \frac{d^3k}{(2\pi)^3}\left[ |\chi_k^2(t)| -\frac{1}{2\omega_k(t)}   \right] \nonumber \\
                                                 &\equiv&  \langle\chi^2(t,{\bf x})\rangle - \delta_M
\end{eqnarray}
where $\delta_M$ is the contribution from the Coleman-Weinberg potential.  This proves that our prescription reproduces the one used in \cite{B}.

Having established a scheme for remormalizing the two-point function of the gaussian field $\chi$ it is straightforward to consider higher order correlation functions.  We simply re-write
the four point function as a product of two-point functions using Wick's theorem.  Then each Wick contraction is renormalized as above.  Applying this prescription to (\ref{pwr_phi}) amounts to
\begin{eqnarray}
  \langle \xi^\phi_{k_1}(t)\xi^\phi_{k_2}(t) \rangle_{\mathrm{ren}} &=& \frac{2 g^4 \dot{\phi}^2}{(2\pi)^3}\frac{1}{\Omega_{k_1}^2} \delta^{(3)}({\bf k_1}+{\bf k_2}) \nonumber \\
                                                     &&\hspace{-28mm}\times \int dt'dt''t't''\sin\left[\Omega_{k_1}(t-t')\right]\sin\left[\Omega_{k_1}(t-t'')\right] \nonumber \\
                                                     &&\hspace{-28mm} \times \int d^3k' \left[ \chi_{k_1-k'}(t')\chi_{k_1-k'}^\star(t'') - f_{k_1-k'}(t')f_{k_1-k'}^\star(t'')\right] \nonumber \\
                                                     &&\hspace{-17mm} \times \left[\chi_{k'}(t')\chi_{k'}^\star(t'')- f_{k'}(t')f_{k'}^\star(t'')\right]  \label{pwr_phi_ren}
\end{eqnarray}
%\begin{widetext}
%\begin{eqnarray}
%  \langle \xi^\phi_{k_1}(t)\xi^\phi_{k_2}(t) \rangle_{\mathrm{ren}} &=& \frac{2 g^4 v^2}{(2\pi)^3}\frac{1}{\Omega_{k_1}^2} \delta^{(3)}({\bf k_1}+{\bf k_2}) \, \int dt'dt''t't''\sin\left[\Omega_{k_1}(t-t')\right]\sin\left[\Omega_{k_1}(t-t'')\right] \nonumber \\
%                                  &&\, \int d^3k' \left[ \chi_{k_1-k'}(t')\chi_{k_1-k'}^\star(t'') - f_{k_1-k'}(t')f_{k_1-k'}^\star(t'') \right] \, \left[ \chi_{k'}(t')\chi_{k'}^\star(t'') - f_{k'}(t')f_{k'}^\star(t'') \right]\label{pwr_phi_ren}
%\end{eqnarray}
%\end{widetext}
where $f_k(t)$ are the adiabatic modes defined in (\ref{f_k}).

\subsection{Spectrum of Re-scattered Modes}

Let us now proceed to compute analytically the renormalized spectrum $P_\phi$ of re-scattered inflaton modes by inserting the solutions (\ref{chi_out}) and (\ref{f_k}) into (\ref{pwr_phi_ren}) 
and carrying out the integrations.  The computation is tedious but straightforward since the time and phase space integrals factorize.  We have relegated the technical details to appendix B and here we simply
state the final result:
\begin{widetext}
\begin{eqnarray}
  P_\phi &=& \frac{g^2}{16\pi^5}\frac{k^3\, k_\star}{k^2 + m^2} \left[ \,\,\,\,\,\,\,\, \frac{e^{-\pi k^2 / (2k_\star^2)}}{2\sqrt{2}}\left( \frac{\pi}{4}|F|^2 + \frac{k_\star^2}{\Omega_k^2}\left[1-\cos(\Omega_k t)\right]^2    \right)   \right. \nonumber \\
 &+& \left[ e^{-\pi k^2/(4k_\star^2)} + \frac{1}{2\sqrt{2}}e^{-3\pi k^2/(8k_\star^2)} \right]\left( -\frac{\pi}{4}\mathrm{Re}\left[ e^{2i\Omega_k t - i\Omega_k^2 / (2k_\star^2) - i\pi/2} F \right] 
                                                                                                     +  \frac{k_\star^2}{\Omega_k^2}\left[1-\cos(\Omega_k t)\right]^2  \right)   \nonumber \\
  &+& \left.  \left[ \frac{4\sqrt{2}}{3\sqrt{3}} e^{-\pi k^2 / (3k_\star^2)} + \frac{2\sqrt{2}}{5\sqrt{5}} e^{-3 \pi k^2/(5k_\star^2)} \right] \frac{\sqrt{\pi}\, k_\star}{\Omega_k} 
              \left[1-\cos(\Omega_k t)\right] \mathrm{Im}\left[ e^{i\Omega_k t - i\Omega_k^2 / (4k_\star^2) - i\pi/4} F \right]\,\,\,\,\,\,\,\, \right] \label{full_pwr_result}
%  && \,\,\,\, \left. \times \frac{\sqrt{\pi} k_\star}{2 \Omega_k}\left[1-\cos(\Omega_kt)\right]\mathrm{Re}\left[e^{i\Omega_kt-i\Omega_k^2 / (4k_\star^2) - i\pi/4}\, F(k,t) \right] \,\,\,\,  \right] \label{full_pwr_result}
\end{eqnarray}
\end{widetext}
Equation (\ref{full_pwr_result}) is the main result of this appendix.  The ``form factor'' $F(k,t)$ is given explicitly in appendix B.

\renewcommand{\theequation}{B-\arabic{equation}}
\setcounter{equation}{0}
\section*{APPENDIX B: Detailed Computation of $P_\phi$}
\label{appB}

\renewcommand{\thesubsection}{B.\arabic{subsection}}
\setcounter{subsection}{0}

In this appendix we discuss in some detail the technical details associated with the computation of $P_\phi$. 
Inserting the solutions (\ref{chi_out}) and (\ref{f_k}) into (\ref{pwr_phi_ren}) 
we find the result
\begin{widetext}
\begin{eqnarray}
  P_\phi &=& \frac{g^2}{8\pi^5}\frac{k^3}{k^2 + m^2} %\label{full_pwr} \\
   \left[\,\,\,\, \int d^3k' n_{k-k'}n_{k'} \int dt'dt'' \sin\left[\Omega_k(t-t')\right]\sin\left[\Omega_k(t-t'')\right] \cos^2\left[\frac{(k_\star t')^2}{2} - \frac{(k_\star t'')^2}{2}\right]  \right. \nonumber \\
  && + \,\,\int d^3k'\sqrt{n_{k-k'}n_{k'}}\sqrt{1+n_{k-k'}}\sqrt{1+n_{k'}} \nonumber \\
  &&\,\,\,\,\,\, \times \int dt'dt'' \sin\left[\Omega_k(t-t')\right]\sin\left[\Omega_k(t-t'')\right] \sin^2\left[\frac{(k_\star t')^2}{2} + \frac{(k_\star t'')^2}{2}\right]   \nonumber \\
  && + \,\,\int d^3k'\left[ n_{k-k'} \sqrt{n_{k'}}\sqrt{1+n_{k'}}  +  n_{k'} \sqrt{n_{k-k'}}\sqrt{1+n_{k-k'}} \right] \nonumber \\
  && \,\,\,\,\,\, \times \left. \int dt'dt'' \sin\left[\Omega_k(t-t')\right]\sin\left[\Omega_k(t-t'')\right] \sin\left[\frac{(k_\star t')^2}{2} + \frac{(k_\star t'')^2}{2}\right]\cos\left[\frac{(k_\star t')^2}{2} - \frac{(k_\star t'')^2}{2}\right]\,\,\,\, 
     \right] \label{full_pwr}
\end{eqnarray}
\end{widetext}
We consider the time and phase space integrations separately.

\subsection{Time Integrals}

\begin{figure}[htbp]
\bigskip \centerline{\epsfxsize=0.35\textwidth\epsfbox{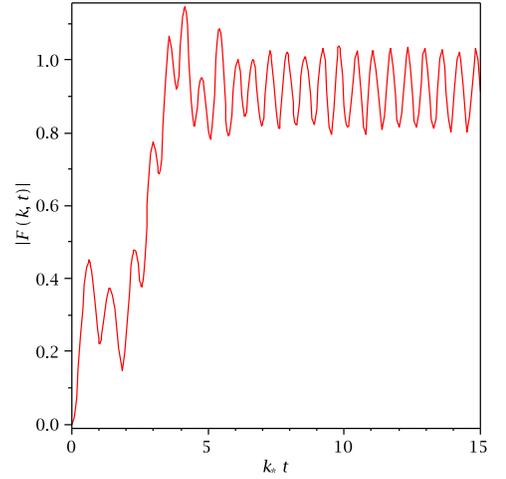}}
%\begin{verse}
%\vskip-0.25cm
\caption{The behaviour of the function $F(k,t)$ as a function of $t$.  For illustration we have set $\Omega_k = 5 k_\star$.
}
\label{Fig:F}
\end{figure}

All the time integrals appearing in (\ref{full_pwr}) can be written in terms of two functions which we call $I_1$, $I_2$.  These involves are defined as
\begin{eqnarray}
  I_1(k,t) &=& \int_0^tdt'\sin\left[\Omega_k(t-t')\right] e^{i(k_\star t')^2} \label{I1_def} \\
  I_2(k,t) &=& \int_0^tdt'\sin\left[\Omega_k(t-t')\right]  \label{I2_def}
\end{eqnarray}
First consider $I_1$.  It is useful to factorize the answer into the product of the stationary phase result (valid for $k_\star t \gg \Omega_k / (2k_\star) \gg 1$) 
and a ``form factor'' $F(k,t)$ as follows:
\begin{eqnarray}
  I_1(k,t) &=& \frac{\sqrt{\pi}}{2k_\star} e^{i\Omega_k t - i\Omega_k^2 / (4k_\star^2) - i\pi / 4} F(k,t) \label{I1}\\
  F(k,t) &=& \frac{1}{2}\left[ \left(1+e^{-2i\Omega_k t}\right)\mathrm{erf}\left(\frac{e^{-i\pi/4}}{2}\frac{\Omega_k}{k_\star}\right) \right. \label{F} \\
  &&\,\,\,\,\, -  \mathrm{erf}\left(\frac{e^{-i\pi/4}}{2}\left(\frac{\Omega_k}{k_\star} - 2k_\star t\right)\right)  \nonumber \\
  &&\,\,\,\,\, \left. -  e^{-2i\Omega_k t}\,\mathrm{erf}\left(\frac{e^{-i\pi/4}}{2}\left(\frac{\Omega_k}{k_\star} + 2k_\star t\right)\right) \right] \nonumber
\end{eqnarray}
The form factor $F(k,t)$ has a complicated structure.  We have illustrated the qualitative behaviour of this function in Fig.\ \ref{Fig:F} taking $\Omega_k / k_\star = 5$ for illustration.

Next, consider the characteristic integral $I_2$, eqn.\ (\ref{I2}).  This integration is trivial:
\begin{equation}
\label{I2}
  I_2(k,t) = \frac{1}{\Omega_k}\left[1-\cos(\Omega_k t)\right]
\end{equation}

Now we will show that all the time integrals appearing in (\ref{full_pwr}) can be reduced to combinations of the characteristic functions $I_1$ and $I_2$. 
First, consider the first line of (\ref{full_pwr}) where the following integral appears:
\begin{eqnarray}
&& \int dt'dt'' \sin\left[\Omega_k(t-t')\right]\sin\left[\Omega_k(t-t'')\right] \nonumber \\
&&\,\,\,\,\,\,\,\,\times \cos^2\left[\frac{(k_\star t')^2}{2} - \frac{(k_\star t'')^2}{2}\right] \nonumber \\
&&\,\,\,\,= \frac{|I_1(k,t)|^2}{2} + \frac{I_2(k,t)^2}{2} \nonumber \\
&&\,\,\,\,= \frac{\pi}{8k_\star^2}|F(k,t)|^2 + \frac{1}{2\Omega_k^2}\left[1-\cos(\Omega_kt)\right]^2
\end{eqnarray}

Next, consider the time integration on the second line of (\ref{full_pwr}):
\begin{eqnarray}
&&  \int dt'dt'' \sin\left[\Omega_k(t-t')\right]\sin\left[\Omega_k(t-t'')\right] \nonumber \\
&&\,\,\,\,\,\,\,\,\times \sin^2\left[\frac{(k_\star t')^2}{2} + \frac{(k_\star t'')^2}{2}\right] \nonumber \\
&& \,\,\,\, = -\frac{\mathrm{Re}\left[I_1(k,t)^2\right]}{2} + \frac{I_2(k,t)^2}{2} \nonumber \\
&& \,\,\,\, = -\frac{\pi}{8k_\star^2}\mathrm{Re}\left[e^{2i\Omega_kt-i\Omega_k^2 / (2k_\star^2) - i\pi/2}\, F(k,t)^2 \right] \nonumber \\
&& \,\,\,\,\,\,\,\,\,\,\,\, + \frac{1}{2\Omega_k^2}\left[1-\cos(\Omega_kt)\right]^2
\end{eqnarray}

Finally, we consider the time integration on the third line of (\ref{full_pwr}):
\begin{eqnarray}
&&\int dt'dt'' \sin\left[\Omega_k(t-t')\right]\sin\left[\Omega_k(t-t'')\right] \\
&&\,\,\,\,\,\,\,\,\times\sin\left[\frac{(k_\star t')^2}{2} + \frac{(k_\star t'')^2}{2}\right]\cos\left[\frac{(k_\star t')^2}{2} - \frac{(k_\star t'')^2}{2}\right] \nonumber \\
&&\,\,\,\, = \mathrm{Im}\left[I_1(k,t) I_2(k,t)\right] \nonumber \\
&&\,\,\,\, = \frac{\sqrt{\pi}}{2k_\star \Omega_k}\left[1-\cos(\Omega_kt)\right]\mathrm{Im}\left[e^{i\Omega_kt-i\Omega_k^2 / (4k_\star^2) - i\pi/4}\, F(k,t) \right] \nonumber
\end{eqnarray}

\subsection{Phase Space Integrals}

Throughout the calculation integrals of the following form appears frequently:
\begin{eqnarray}
   && \int d^3k' n_{k-k'}^a n_{k'}^b \nonumber \\
  && = \int d^3k' \exp\left[-a\pi |{\bf k} - {\bf k'}|^2 / k_\star^2 \right]\exp\left[-b\pi |{\bf k'}|^2 / k_\star^2 \right]  \nonumber \\
   && =  \frac{k_\star^3}{(a+b)^{3/2}} \exp\left[-\frac{ab}{a+b}\frac{\pi k^2}{k_\star^2}\right] \label{phase_identity}
\end{eqnarray}
This formula is valid when $a$, $b$ are positive real numbers. Notice that this expression is symmetric under interchange of $a$ and $b$.

The phase space integral in the first line of (\ref{full_pwr}) is computed by a trivial application of the identity (\ref{phase_identity}):
\begin{equation}
 \int d^3k' n_{k-k'}n_{k'} =  \frac{k_\star^3}{2\sqrt{2}} e^{-\pi k^2 / (2 k_\star^2)}
\end{equation}

The remaining integrals cannot be obtained exactly in closed form because they contain terms like $\sqrt{1+n_{k'}}$ where the gaussian factors appear under the square root.  However, 
because $n_k \leq 1$ it turns out to be a very good approximation to replace  $\sqrt{1+n_{k'}} \cong 1 + n_{k'}/2$.  (We have checked numerically that the error induced is less than a few percent.)
Let us now proceed in this manner.  The phase space integral on the second line of (\ref{full_pwr}) is:
\begin{eqnarray}
 && \int d^3k' \sqrt{n_{k-k'}n_{k'}}\sqrt{1+n_{k-k'}}\sqrt{1+n_{k'}} \nonumber \\
 && \cong \int d^3k'\left[  n_{k-k'}^{1/2}n_{k'}^{1/2}  + \frac{1}{2} n_{k-k'}^{3/2}n_{k'}^{1/2} + \frac{1}{2} n_{k-k'}^{1/2}n_{k'}^{3/2}  \right] \nonumber \\
 && = k_\star^3 \left[ \exp\left(-\frac{\pi k^2}{4k_\star^2}\right) + \frac{1}{2\sqrt{2}}\exp\left(-\frac{3 \pi k^2}{8k_\star^2}\right) \right]
\end{eqnarray}

Finally, consider the phase space integral on the third line of (\ref{full_pwr}):
\begin{eqnarray}
 && \int d^3k' \left[ n_{k-k'} \sqrt{n_{k'}}\sqrt{1+n_{k'}}  +  n_{k'} \sqrt{n_{k-k'}}\sqrt{1+n_{k-k'}}  \right] \nonumber \\
 && \cong \int d^3k' \left[  n_{k-k'}n_{k'}^{1/2} + n_{k'}n_{k-k'}^{1/2}  \right. \nonumber \\
 && \,\,\,\,\,\,\,\, \hspace{15mm}\left. + \frac{1}{2} n_{k-k'}n_{k'}^{3/2}  + \frac{1}{2} n_{k'}n_{k-k'}^{3/2}       \right] \nonumber \\
 && = k_\star^3 \left[\frac{4\sqrt{2}}{3\sqrt{3}} \exp\left(-\frac{\pi k^2}{3k_\star^2}\right) + \frac{2\sqrt{2}}{5\sqrt{5}} \exp\left(-\frac{3 \pi k^2}{5k_\star^2}\right) \right]
\end{eqnarray}

Assembling the various results presented in this appendix one arrives straightforwardly at the result (\ref{full_pwr_result}).


\begin{thebibliography}{999}

\bibitem{fluct}
 V.~F.~Mukhanov and G.~V.~Chibisov,
%``Quantum Fluctuation And 'Nonsingular' Universe,''
JETP Lett.\  {\bf 33}, 532 (1981)
[Pisma Zh.\ Eksp.\ Teor.\ Fiz.\  {\bf 33}, 549 (1981)]; 
S.~W.~Hawking,
%``The Development Of Irregularities In A Single Bubble Inflationary Universe,''
Phys.\ Lett.\ B {\bf 115}, 295 (1982); 
A.~A.~Starobinsky,
%``Dynamics Of Phase Transition In The New Inflationary Universe Scenario And Gen
%eration Of Perturbations,''
Phys.\ Lett.\ B {\bf 117}, 175 (1982); 
A.~H.~Guth and S.~Y.~Pi,
%``Fluctuations In The New Inflationary Universe,''
Phys.\ Rev.\ Lett.\  {\bf 49}, 1110 (1982); J.~M.~Bardeen, P.~J.~Steinhardt and
M.~S.~Turner,
%``Spontaneous Creation Of Almost Scale - Free Density Perturbations In An Inflationary Universe,''
Phys.\ Rev.\ D {\bf 28}, 679 (1983).


\bibitem{modulated}

  L.~Kofman,
  %``Probing string theory with modulated cosmological fluctuations,''
  arXiv:astro-ph/0303614.
  %%CITATION = ASTRO-PH/0303614;%%

  F.~Bernardeau, L.~Kofman and J.~P.~Uzan,
  %``Modulated fluctuations from hybrid inflation,''
  Phys.\ Rev.\  D {\bf 70}, 083004 (2004)
  [arXiv:astro-ph/0403315].
  %%CITATION = PHRVA,D70,083004;%%

\bibitem{modulated2}

   G.~Dvali, A.~Gruzinov and M.~Zaldarriaga,
  %``A new mechanism for generating density perturbations from inflation,''
  Phys.\ Rev.\  D {\bf 69}, 023505 (2004)
  [arXiv:astro-ph/0303591].
  %%CITATION = PHRVA,D69,023505;%%

  G.~Dvali, A.~Gruzinov and M.~Zaldarriaga,
  %``Cosmological perturbations from inhomogeneous reheating, freezeout, and
  %mass domination,''
  Phys.\ Rev.\  D {\bf 69}, 083505 (2004)
  [arXiv:astro-ph/0305548].
  %%CITATION = PHRVA,D69,083505;%%

\bibitem{curvaton}

  D.~H.~Lyth and D.~Wands,
  %``Generating the curvature perturbation without an inflaton,''
  Phys.\ Lett.\  B {\bf 524}, 5 (2002)
  [arXiv:hep-ph/0110002].
  %%CITATION = PHLTA,B524,5;%%



\bibitem{KL}

  L.~A.~Kofman and A.~D.~Linde,
 % ``Generation of Density Perturbations in the Inflationary Cosmology,''
  Nucl.\ Phys.\  B {\bf 282}, 555 (1987).
  %%CITATION = NUPHA,B282,555;%%

\bibitem{KP}

  L.~A.~Kofman and D.~Y.~Pogosian,
 % ``NONFLAT PERTURBATIONS IN INFLATIONARY COSMOLOGY,''
  Phys.\ Lett.\  B {\bf 214}, 508 (1988).
  %%CITATION = PHLTA,B214,508;%%


\bibitem{BBS}

 D.~S.~Salopek, J.~R.~Bond and J.~M.~Bardeen,
 % ``Designing Density Fluctuation Spectra in Inflation,''
  Phys.\ Rev.\  D {\bf 40}, 1753 (1989).
  %%CITATION = PHRVA,D40,1753;%%

\bibitem{ng}

 N.~Barnaby and J.~M.~Cline,
  %``Nongaussian and nonscale-invariant perturbations from tachyonic  preheating
  %in hybrid inflation,''
  Phys.\ Rev.\  D {\bf 73}, 106012 (2006)
  [arXiv:astro-ph/0601481].
  %%CITATION = PHRVA,D73,106012;%%

  N.~Barnaby and J.~M.~Cline,
  %``Nongaussianity from Tachyonic Preheating in Hybrid Inflation,''
  Phys.\ Rev.\  D {\bf 75}, 086004 (2007)
  [arXiv:astro-ph/0611750].
  %%CITATION = PHRVA,D75,086004;%%

\bibitem{T}

  D.~J.~H.~Chung, E.~W.~Kolb, A.~Riotto and I.~I.~Tkachev,
 % ``Probing Planckian physics: Resonant production of particles during
 % inflation and features in the primordial power spectrum,''
  Phys.\ Rev.\  D {\bf 62}, 043508 (2000)
  [arXiv:hep-ph/9910437].
  %%CITATION = PHRVA,D62,043508;%%

\bibitem{B}

  L.~Kofman, A.~Linde, X.~Liu, A.~Maloney, L.~McAllister and E.~Silverstein,
  %``Beauty is attractive: Moduli trapping at enhanced symmetry points,''
  JHEP {\bf 0405}, 030 (2004)
  [arXiv:hep-th/0403001].
  %%CITATION = JHEPA,0405,030;%%

\bibitem{S}

 A.~E.~Romano and M.~Sasaki,
 % ``Effects of particle production during inflation,''
  arXiv:0809.5142 [gr-qc].
  %%CITATION = ARXIV:0809.5142;%%

\bibitem{E}

  D.~Green, B.~Horn, L.~Senatore and E.~Silverstein,
  %``Trapped Inflation,''
  arXiv:0902.1006 [hep-th].
  %%CITATION = ARXIV:0902.1006;%%

\bibitem{KL99} 

L.~A.~Kofman and A.~D.~Linde, ``Trapped Inflation,'' unpublished (1999).

\bibitem{E0}

  E.~Silverstein and A.~Westphal,
 % ``Monodromy in the CMB: Gravity Waves and String Inflation,''
  Phys.\ Rev.\  D {\bf 78}, 106003 (2008)
  [arXiv:0803.3085 [hep-th]].
  %%CITATION = PHRVA,D78,106003;%%

\bibitem{KLS}

 L.~Kofman, A.~D.~Linde and A.~A.~Starobinsky,
 % ``Reheating after inflation,''
  Phys.\ Rev.\ Lett.\  {\bf 73}, 3195 (1994)
  [arXiv:hep-th/9405187].
  %%CITATION = PRLTA,73,3195;%%

\bibitem{KLS97}

  L.~Kofman, A.~D.~Linde and A.~A.~Starobinsky,
 %``Towards the theory of reheating after inflation,''
  Phys.\ Rev.\  D {\bf 56}, 3258 (1997)
  [arXiv:hep-ph/9704452].
  %%CITATION = PHRVA,D56,3258;%%

\bibitem{SUSY}

  A.~Berera and T.~W.~Kephart,
  %``The interaction structure and cosmological relevance of mass scales in
  %string motivated supersymmetric theories,''
  Phys.\ Lett.\  B {\bf 456}, 135 (1999)
  [arXiv:hep-ph/9811295].
  %%CITATION = PHLTA,B456,135;%%

\bibitem{FT}

  G.~N.~Felder and I.~Tkachev,
%  ``LATTICEEASY: A program for lattice simulations of scalar fields in an
%  expanding universe,''
  [arXiv:hep-ph/0011159].
  %%CITATION = CPHCB,178,;%%

\bibitem{Fr}

  A.~V.~Frolov,
%  ``DEFROST: A New Code for Simulating Preheating after Inflation,''
  JCAP {\bf 0811}, 009 (2008)
  [arXiv:0809.4904 [hep-ph]].
  %%CITATION = JCAPA,0811,009;%%

\bibitem{TM}

  R.~Micha and I.~I.~Tkachev,
  %``Relativistic turbulence: A long way from preheating to equilibrium,''
  Phys.\ Rev.\ Lett.\  {\bf 90}, 121301 (2003)
  [arXiv:hep-ph/0210202].
  %%CITATION = PRLTA,90,121301;%%

\bibitem{MT}

  R.~Micha and I.~I.~Tkachev,
  %``Turbulent thermalization,''
  Phys.\ Rev.\  D {\bf 70}, 043538 (2004)
  [arXiv:hep-ph/0403101].
  %%CITATION = PHRVA,D70,043538;%%

\bibitem{B1}

  J.~Berges, A.~Rothkopf and J.~Schmidt,
  %``Non-thermal fixed points: effective weak-coupling for strongly correlated
  %systems far from equilibrium,''
  Phys.\ Rev.\ Lett.\  {\bf 101}, 041603 (2008)
  [arXiv:0803.0131 [hep-ph]].
  %%CITATION = PRLTA,101,041603;%%

\bibitem{B2}

  J.~Berges and G.~Hoffmeister,
  %``Nonthermal fixed points and the functional renormalization group,''
  arXiv:0809.5208 [hep-th].
  %%CITATION = ARXIV:0809.5208;%%

\bibitem{inst}

G.~Felder, L.~Kofman and A.~Linde,
Phys.\ Rev.\  {\bf D59}, 123523 (1999)
[hep-ph/9812289].

\bibitem{lattice} S. Khlebnikov and I.~I.~Tkachev, Phys. Rev. Lett.
{\bf 77}, 219 (1996); Phys. Rev. Lett. {\bf 79}, 1607 (1997);
Phys. Rev. {\bf D56}, 653 (1997).

\bibitem{FK}
  G.~N.~Felder and L.~Kofman,
  %``The development of equilibrium after preheating,''
  Phys.\ Rev.\  D {\bf 63}, 103503 (2001)
  [arXiv:hep-ph/0011160].
  %%CITATION = PHRVA,D63,103503;%%

\bibitem{spikes}

  J.~R.~Bond, A.~V.~Frolov, Z.~Huang and L.~Kofman,
  %``Non-Gaussian Spikes from Chaotic Billiards in Inflation Preheating,''
  arXiv:0903.3407 [astro-ph.CO].
  %%CITATION = ARXIV:0903.3407;%%

\bibitem{neilprog}

N.\ Barnaby, in preparation.

\bibitem{in-in}

J.\ Schwinger, Proc.\ Nat.\ Acad.\ Sci.\ US {\bf 46}, 1401 (1961).

\bibitem{weinberg}

  S.~Weinberg,
  %``Quantum contributions to cosmological correlations,''
  Phys.\ Rev.\  D {\bf 72}, 043514 (2005)
  [arXiv:hep-th/0506236].
  %%CITATION = PHRVA,D72,043514;%%

\bibitem{seery}

  D.~Seery, K.~A.~Malik and D.~H.~Lyth,
  %``Non-gaussianity of inflationary field perturbations from the field
  %equation,''
  JCAP {\bf 0803}, 014 (2008)
  [arXiv:0802.0588 [astro-ph]].
  %%CITATION = JCAPA,0803,014;%%

\bibitem{tree}

  S.~Weinberg,
  %``A Tree Theorem for Inflation,''
  arXiv:0805.3781 [hep-th].
  %%CITATION = ARXIV:0805.3781;%%



\end{thebibliography}
\end{document}